# The ρ-Geminid meteoroid stream: orbits, spectroscopic data and implications for its parent body

José M. Madiedo[1, 2]

[1] Facultad de Ciencias Experimentales, Universidad de Huelva. 21071 Huelva (Spain).

[2] Facultad de Física, Universidad de Sevilla, Departamento de Física Atómica, Molecular y Nuclear, 41012 Sevilla, Spain

**ABSTRACT**

By using an array of high-sensitivity CCD video devices and spectrographs, the activity of meteor events from the poorly-known ρ-Geminid meteoroid stream has been monitored during January 2012 and 2013. As a result of this research, the atmospheric trajectory and radiant position of 10 of these events has been obtained, but also the orbital parameters of the progenitor meteoroids and the tensile strength of these particles. The data reveal that the progenitor of this stream must be a comet. In addition, the emission spectra produced by three ρ-Geminid meteors were also recorded. These are the first ρ-Geminid spectra discussed in the scientific literature, and have provided clues about the chemical nature of these meteoroids and their parent body.

**KEYWORDS:** meteorites, meteoroids, meteors.

**1 INTRODUCTION**

The ρ-Geminid meteor shower was discovered by Southworth and Hawkins (1963) in the framework of the 1952-1954 Harvard Meteor Program. This poorly-known stream is currently included in the IAU working list of meteor showers with the code 94 RGE (http://www.astro.amu.edu.pl/~jopek/MDC2007/).  Its activity extends from December 28 to January 28, peaking around January 8, and its parent body remains unknown. Data about this shower are scarce (Southworth and Hawkins 1963; Lindblad 1971a,b; Sekanina 1973), mainly because of its low zenithal hourly rate (ZHR) and typical bad weather conditions in the northern hemisphere during the above-mentioned activity period. This ZHR was of around 12 h$^{-1}$ during the outburst exhibited on January 16-17, 1993 (Jenniskens 1996). In order to get a better knowledge about the ρ-Geminid meteoroid stream, additional observations of the RGE shower providing accurate enough orbital data would be necessary. With this aim, a systematic monitoring of meteor activity has been conducted in January 2012 and 2013 by using an array of high-sensitivity video devices. These devices have been proven to be very useful to obtain orbital data of meteoroids ablating in the atmosphere, and, in fact, these have become very popular in recent years (Madiedo & Trigo-Rodríguez 2008, Jenniskens et al. 2011). This paper focuses on the analysis of the multistation ρ-Geminid meteors imaged during this monitoring. Thus, atmospheric trajectories, radiant position and orbital data are presented, and several asteroids are identified as potential progenitors of this stream. In addition, three emission spectra produced by bright RGE meteors are presented and discussed. These are, to my knowledge, the first ρ-Geminid spectra discussed in the scientific literature. These spectra have provided new clues about the chemical nature of these meteoroids and their parent body.

## 2 INSTRUMENTATION AND METHODS

To image the ρ-Geminids discussed in this work a set of high-sensitivity CCD video devices operating at the meteor observing stations listed in Table 1 were employed. Each station operates an array of 4 to 12 Watec video cameras (models 902H and 902H2 Ultimate) to monitor the night sky (Madiedo & Trigo-Rodríguez 2008). These CCD video cameras generate interlaced imagery following the PAL video standard at a rate of 25 frames per second (fps) and with a resolution of 720x576 pixels. Their focal length ranges from 6 to 25 mm and the field of view covered by each device ranges, approximately, from 62x50 to 14x11 degrees. To get an insight about the chemical nature of the progenitor meteoroids, attached holographic diffraction gratings (with 500 or 1000 lines/mm, depending on the device) were attached to the objective lens of some of these CCD cameras. These videospectrographs can record the emission spectra produced by meteor events with luminosity higher than mag. -3/-4. The meteor stations work in a fully autonomous way by means of the MetControl software (Madiedo et al. 2010). A more detailed description of the operation of these systems is given in (Madiedo & Trigo-Rodríguez 2008; Madiedo 2014).

For data reduction, an astrometric measurement is done by hand in order to obtain the plate (x, y) coordinates of the meteor along its apparent path from each observing station. The astrometric measurements are then introduced in the AMALTHEA software (Madiedo et al. 2011), which transforms plate coordinates into equatorial coordinates by using the position of reference stars appearing in the images. Typically, over 40 reference stars are used. This package employs the method of the intersection of planes to determine the position of the apparent radiant and also to reconstruct the trajectory in the atmosphere of meteors recorded from at least two different observing stations (Ceplecha 1987). The preatmospheric velocity $V_\infty$ is found by measuring the velocities at the earliest part of the meteor trajectory. Once these data are known, the software computes the orbital elements of the parent meteoroid by following the method described in Ceplecha (1987).

**3 OBSERVATIONS AND RESULTS**

Weather conditions in Spain during January 2012 and 2013 posed important difficulties to the monitoring of the ρ-Geminids during most of its activity period. In spite of this, 21 multi-station RGE meteors were imaged from the meteor observing stations listed in Table 1 from January 1 to January 18. Those events with a convergence angle below 20º were not taken into account. This is the angle between the two planes delimited by the observing stations and the meteor path in the triangulation, and it measures the quality of the determination of the atmospheric trajectory (Ceplecha 1987). By following this quality criterion 11 RGE meteors were discarded. So, as a result of the above-mentioned monitoring campaign, 10 double-station RGE meteors were considered for this research. Two of these were bright enough to produce emission spectra that were recorded by the videospectrographs. The recording dates and times of these 10 events are listed in Table 2, where their assigned SPMN codes are used for identification. The analysis of the video images provided their atmospheric trajectory, and the main parameters of these paths are also listed in Table 2 together with the position of the geocentric radiant (J2000) and the absolute magnitude of each event. The average value of the observed preatmospheric velocity of these particles yields $V_\infty = 22.9 \pm 0.4$ km s$^{-1}$. The photometric preatmospheric mass $m_p$ of each meteoroid is also included in this Table. As in previous papers (see e.g. Madiedo et al. 2013a), this mass has been calculated from the lightcurve of each event and by using the luminous efficiency given by Ceplecha and McCrosky (1976). This mass was found to range between 0.2 and 100 g.

### 3.1 Orbital data

Table 3 shows the orbital elements obtained for the progenitor meteoroids of the meteor events listed in Table 2, and also the averaged orbital data calculated by taking into account these N=10 orbits. In Table 3 a is the semi-major axis, e the orbital eccentricity, i the inclination, ω the argument of perihelion, Ω the longitude of the ascending node and q the perihelion distance. The values of the orbital period P and the Tisserand parameter with respect to Jupiter ($T_J$) are also listed. To validate the association of these events with the ρ-Geminid stream, the Southworth & Hawkins $D_{SH}$ dissimilarity function (Southworth & Hawkins 1963) was calculated. As Table 2 shows, $D_{SH}$ values obtained by comparing the orbit of this stream (Lindblad 1971a, Jenniskens 2006) with the orbits of these meteoroids remain below the usual cut-off value of 0.15 adopted to establish a valid link (Lindblad 1971a, 1971b).

### 3.2 Emission spectra

The emission spectra of three ρ-Geminid fireballs (events with SPMN code 060112, 130113 and 180113 in Table 2) were imaged by the videospectrographs. However, those of meteors SPMN060112 and SPMN180113 were of very limited use, since the signal was not bright enough to show contributions other than the corresponding to the Na doublet at 588.9 nm and the Mg triplet at 516.7 nm. Fortunately the third spectrum, which was produced by event SPMN130113, showed much more contributions. These spectra were reduced by following the standard procedure described in Madiedo et al. (2013b). Thus, the video files containing these signals were deinterlaced, sky-background-subtracted and flat-fielded. Next, each spectrum was calibrated in wavelength by using typical metal lines appearing in meteor spectra, and then corrected by taking into account the instrumental efficiency of the recording device.

The calibrated spectrum of fireball SPMN130113 is shown in Figure 1, where the most important contributions have been highlighted and multiplet numbers are given according to Moore (1945). The most significant contributions correspond to the Mg I-2 triplet at 516.7 nm and the Na I-1 doublet at 588.9 nm. The emission from Ca I-2 (422.6 nm), Fe I-41 (441.5 nm) and Fe I-15 (526.9 nm) are also very noticeable, together with the signal from Fe I-4 (385.6 nm) and Mg I-3 (383.8 nm), although these two are not individually resolved. In addition, the emission from FeO was identified around the Na I-1 line.

## 4 DISCUSSION

### 4.1 Meteoroid strength

According to Ceplecha (1988), meteoroids can be classified into four different populations or groups on the basis of the so-called $K_B$ parameter. This parameter is given by the following equation:

$$K_B = \log \rho_B + 2.5 \log V_\infty - 0.5 \log \cos z_R + 0.15 \qquad (1)$$

where $\rho_B$ is the air density at the beginning of the luminous trajectory (in g cm$^{-3}$), $V_\infty$ is the preatmospheric velocity of the meteoroid (in cm s$^{-1}$), and $z_R$ is the inclination of the atmospheric trajectory with respect to the vertical. Thus, depending on the value of $K_B$, meteoroids can belong to any of the following groups: A-group, comprising particles similar to carbonaceous chondrites ($7.3 \leq K_B < 8$); B-group of dense cometary material ($7.1 \leq K_B < 7.3$); C group of regular cometary material ($6.6 \leq K_B < 7.1$); and D-group of soft cometary material ($K_B < 6.6$). In this work the air density $\rho_B$ in equation (1) was obtained from the US standard atmosphere model (U.S. Standard Atmosphere 1976). The calculation for the ρ-Geminids listed in Table 2 yields $K_B = 7.4 \pm 0.3$. So, within the experimental uncertainty of the data analyzed here, it is not clear weather RGE meteoroids would belong to the population of dense cometary material (group B) or to the population of particles similar to carbonaceous chondrites (group A). In any event, this value of $K_B$ suggests that these particles are tougher than regular cometary material.

The PE parameter (Ceplecha 1988), which has been traditionally employed to compare the strengths of large meteoroids, was also calculated. It is given by the following equation:

$$PE = \log \rho_E - 0.42 \log m_\infty + 1.49 \log V_\infty - 1.29 \log \cos z_R \qquad (2)$$

where $\rho_E$ is the air density at the terminal point of the luminous trajectory. The value of PE classifies meteoroids according to the following groups: group I, of ordinary chondritic meteoroids (-4.6 < PE); group II, of early type carbonaceous chondrites (-5.25 < PE ≤ -4.6); and groups IIIa (-5.70 < PE ≤ -5.25) and IIIb (PE ≤ -5.70), of cometary materials. For the ρ-Geminid events analyzed here it was found that PE = -5.02 ± 0.31. This result shows that it is not possible to establish clearly if, from the point of view of the PE parameter, RGE meteoroids belong to the group of carbonaceous chondrites (group II) or to group IIIa of cometary materials.

The toughness of RGE meteoroids can be investigated by estimating their tensile strength. This strength can be calculated by analyzing the flares exhibited by RGE meteors, as in previous papers (see e.g. Madiedo et al. 2013c). Most of the double-station ρ-Geminids in Table 2 exhibited one flare along their atmospheric path. These flares took place during the second half of their luminous trajectory. These flares occur as a consequence of the sudden fragmentation of the meteoroids in the atmosphere when the aerodynamic pressure becomes larger than the meteoroid strength. This pressure P is given by the following equation:

$$P = \rho_{atm} \cdot v^2 \qquad (3)$$

where v is the velocity of the meteoroid and $\rho_{atm}$ the atmospheric density at the corresponding height. The flare is produced by the fast ablation of tiny fragments produced during this fragmentation, which are delivered to the thermal wave in the fireball's bow shock. According to this, the critical pressure under which the fragmentation takes place can be used as an estimation of the tensile strength of the meteoroid (Trigo-Rodriguez & Llorca 2006, 2007).

The analysis of the video images has provided the values summarized in Table 4. This shows the break-up height and velocity for each meteoroid, together with the inferred tensile strength. The atmospheric density in equation (3) was estimated again by employing the US standard atmosphere model (U.S. Standard Atmosphere 1976). This data show that the tensile strength of RGE meteoroids would range from $(6.1 \pm 0.7) \cdot 10^{-2}$ MPa for the 110113 meteor to $(2.1 \pm 0.2) \cdot 10^{-1}$ MPa for the 070113b event, with an average value of $(1.2 \pm 0.1) \cdot 10^{-1}$ MPa. These value are higher that the average strength inferred for meteoroids belonging to old stream, such as the Taurids (~$3.4 \cdot 10^{-2}$ MPa) and the Quadrantids (~2·10$^{-2}$ MPa) (Trigo-Rodríguez & Llorca 2006, 2007). However, they are lower than the (3.0 ± 0.3)·10$^{-1}$MPa strength found for very tough Taurid meteoroids capable to penetrate very deep the atmosphere (Madiedo et al. 2014a). So, the calculated strength values suggest that RGE meteoroids are composed by tough cometary materials and so their parent body must be a comet.

**4.2 Beginning and ending heights**

Figures 2 and 3 show the beginning and terminal heights ($H_b$ and $H_e$, respectively) of meteors listed in Table 2. The only event not included in Figure 3 is the meteor with code 130113, since its ending height was out of the field of view of both recording cameras and so it could not be measured.

The beginning height $H_b$ of ρ-Geminid meteors was found to increase with increasing meteoroid mass. For events listed in Table 2 this height was always below 97 km above the sea level. This behaviour has been also found for other cometary meteor showers, although the beginning heights exhibited by RGE meteors are lower than those exhibited by the Lyrids and the Perseids (Figure 5 in Jenniskens 2004). This can be explained on the basis of the lower preatmospheric velocity of the ρ-Geminids ($V_\infty$ ~23 km s$^{-1}$) with respect to the Lyrids ($V_\infty$ = 49 km s$^{-1}$) and the Perseids ($V_\infty$ = 61 km s$^{-1}$), since the emission of radiation tends to start later for meteoroids moving at lower velocities. The dependence of $H_b$ on the logarithm of the photometric mass was described by means of a linear relationship (solid line in Figure 2). The slope of this straight line yields a = 1.1 ± 0.5. According to this result, the increase of the beginning height with mass is less pronounced for the ρ-Geminid than for other cometary showers such as the Leonids (a = 9.9 ± 1.5), the Perseids (a = 7.9 ± 1.3), the Taurids (a = 6.6 ± 2.2) and the Orionids (a = 5.02 ± 0.65), but more pronounced than for the Geminids (a = 0.46 ± 0.26) (Koten et al. 2004).

Figure 3 shows that, as expected, the terminal point of the luminous trajectory occurs at lower altitudes as the meteoroid mass increases. This behaviour can be described by means of a linear relationship between $H_e$ and the logarithm of the photometric mass (solid line in Figure 3), where the calculated slope of this line is b = -9.3 ± 0.5. The final height of these RGE meteors is similar to the terminal level reached by the Geminids, which have an asteroidal origin (Jenniskens 2004). This does not contradict the result found in the previous section in the sense than tensile strength values suggest that the ρ-Geminid are composed by tough cometary materials. Thus, the lower velocity of RGE meteors (about 23 km s$^{-1}$) with respect to the Geminids (~36 km s$^{-1}$) can explain that both showers exhibit a similar terminal height, since slower meteoroids tend to penetrate deeper in the atmosphere.

### 4.3 Orbital elements and potential parent body

The averaged value of the Tisserand parameter with respect to Jupiter yields $T_J$ = 2.93 ± 0.09. This shows that RGE meteoroids were following a Jupiter family comet orbit before impacting our planet, although this $T_J$ value is very close to the lower limit for typical asteroidal orbits.

Orbital dissimilarity criteria were employed in order to determine the likely parent of the ρ-Geminids. These criteria are based on the calculation of the so-called dissimilarity function, which measured the "distance" between the orbits of the parent candidate and the meteoroid stream. A review about this topic can be found in Williams (2011). According to this approach, a parent body would be associated to a given meteoroid stream if the value of the dissimilarity function falls below an appropriate cut-off value. Nevertheless, a positive similarity at a single time instant is not enough to claim a dynamical association, and it must be tested if this link is just casual or real (Williams 1993, 2004). Thus, the evolution with time of the dissimilarity function must be analyzed in order to confirm that both orbits are similar over a period of time. Porubčan et al. (2004) studied the similarity of meteoroid orbits with their parent bodies for a period of at least 5,000 years before an association is claimed, a view later supported by Trigo-Rodriguez et al. (2007). The ORAS software (Madiedo et al. 2013d) was employed to establish a potential link with one of the NEOs included in the NeoDys database (http://newton.dm.unipi.it/neodys/). This analysis has been performed by calculating the widely employed Southworth and Hawkins $D_{SH}$ dissimilarity function (Southworth & Hawkins 1963), for which the cut-off condition $D_{SH}$ < 0.15 is commonly adopted in order to validate an association (Linblad 1971a,b). The average orbit calculated for RGE meteoroids in Table 3 has been employed for this analysis. Two NEOs have been obtained as parent candidates: 2007VW137 ($D_{SH}$ = 0.09) and 2010AG30 ($D_{SH}$ = 0.13). The evolution with time since the present epoch of the $D_{SH}$ criterion for these objects and the orbit of the ρ-Geminid meteoroid stream is shown in Figure 4. This plot shows that the similarity of the orbits of these NEOs and the ρ-Geminids is lost within a time period of about 350 years in the past for 2007VW137, and 100 years for 2010AG30. So, it is very likely that the link found between these objects and the RGE stream is not real, but casual. As a consequence, from this analysis none of these two objects can be proposed on solid grounds as the potential parent of this stream.

### 4.4. Emission spectra

The SPMN130113 spectrum was studied in order to get an insight into the chemical nature of ρ-Geminid meteoroids. To do so, the emission from multiplets Na I-1, Mg I-2 and Fe I-15 has been analyzed (Borovička et al. 2005). These contributions were measured frame by frame in the corresponding videospectrum, and then corrected by means of the spectral response of the recording device. For each multiplet, these contributions were then added frame by frame to obtain their integrated intensity along the atmospheric path of the fireball. Then, the Na/Mg and Fe/Mg intensity ratios were calculated. From this analysis the ratios Na/Mg = 1.03 and Fe/Mg = 1.57 were obtained. As shown in Figure 5 in Borovička et al. (2005), the Na/Mg intensity ratios fit fairly well the result expected for meteoroids with chondritic composition for a meteor velocity of about 23 km s$^{-1}$. To confirm this result, the Na I-1, Mg I-2 and Fe I-15 relative intensities were plotted in the ternary diagram shown in Figure 5, where the solid curve indicates the expected relative intensity as a function of meteor velocity for chondritic meteoroids (Borovička et al. 2005). As can be seen, the value corresponding to the SPMN130113 emission spectrum fits fairly well the expected relative intensity for a meteor velocity of about 23 km s$^{-1}$, which again suggests a chondritic nature for ρ-Geminid meteoroids.

With respect to the spectra of events SPMN060112 and SPMN180113, which as previously mentioned had a very limited usage, only the Na/Mg intensity ratio could be obtained since they were not bright enough to show the contributions from Fe or any other elements. The values of this intensity ratio were 1.09 and 1.01, respectively. These agree with the Na/Mg intensity ratio obtained for the SPMN130113 spectrum.

## 5 CONCLUSIONS

The activity of the poorly-known ρ-Geminid meteor shower was monitored and analyzed in January 2012 and 2013. As a result of this survey, 10 double-station RGE meteors provided good orbits and atmospheric trajectories. The emission spectra imaged for three of these ρ-Geminid events suggest a chondritic nature for members of this meteoroid stream. Within the experimental uncertainty of the data presented here, the value obtained for the $K_B$ parameter (7.4 ± 0.3) does not clarify weather RGE meteoroids belong to the population of dense cometary materials (group B) or to the population of particles similar to carbonaceous chondrites (group A). However, the calculated value of the Tisserand's parameter shows that this meteoroid stream follows a Jupiter family comet orbit, and the tensile strength estimated for these particles is consistent with tough cometary materials. Besides, the initial height of these meteor events was found to increase as the preatmospheric mass of the meteoroid increases, as has been previously found for cometary meteor showers. According to this, it is concluded that the parent body of this stream is a comet. However, a search among objects currently identified in the NEO population and based on the $D_{SH}$ Southworth and Hawking dissimilarity criterion did not allow finding any candidate as progenitor body.

## ACKNOWLEDGEMENTS

I thank Dr. Jiri Borovička for his valuable comments and suggestions. I am also grateful to AstroHita Foundation for its continuous support in the operation of the meteor observing station located at La Hita Astronomical Observatory.

# **TABLES**

Table 1. Geographical coordinates of the meteor observing stations involved in this research.

| Station # | Station name | Longitude | Latitude (N) | Alt. (m) |
|---|---|---|---|---|
| 1 | Sevilla | 05º 58´ 50" W | 37º 20´ 46" | 28 |
| 2 | La Hita | 03º 11' 00" W | 39º 34' 06" | 674 |
| 3 | Huelva | 06º 56' 11" W | 37º 15' 10" | 25 |
| 4 | Sierra Nevada | 03º 23´ 05" W | 37º 03´ 51" | 2896 |
| 5 | El Arenosillo | 06º 43´ 58" W | 37º 06´ 16" | 40 |

Table 2. Atmospheric trajectory and radiant data (J2000) for the double-station ρ-Geminid meteors analyzed in this work. M: absolute magnitude; $m_p$: photometric mass; $H_b$ and $H_e$: beginning and ending height of the luminous phase, respectively; $\alpha_g, \delta_g$: right ascension and declination of the geocentric radiant; $V_\infty$, $V_g$, $V_h$: observed preatmospheric, geocentric and heliocentric velocities, respectively.

| Meteor code | Date and Time (UTC) ±0.1s | M ±0.5 | $m_p$ | $H_b$ (km) | $H_e$ (km) | $\alpha_g$ (º) | $\delta_g$ (º) | $V_\infty$ (km s$^{-1}$) | $V_g$ (km s$^{-1}$) | $V_h$ (km s$^{-1}$) |
|---|---|---|---|---|---|---|---|---|---|---|
| 030112 | 3 Jan. 2012 2h24m52.8s | -4.5 | 15±2 | 92.6 | 60.0 | 98.68±0.16 | 33.46±0.07 | 23.5±0.5 | 20.9±0.5 | 38.1±0.5 |
| 060112 | 6 Jan. 2012 0h35m16.9s | -3.0 | 6.6±0.6 | 94.5 | 53.1 | 97.98±0.10 | 28.76±0.05 | 22.8±0.3 | 20.0±0.3 | 38.5±0.3 |
| 050113 | 5 Jan. 2013 23h31m38.1s | 0.0 | 0.2±0.02 | 91.8 | 78.5 | 101.47±0.11 | 27.42±0.08 | 23.3±0.4 | 20.4±0.4 | 38.1±0.4 |
| 070113a | 7 Jan. 2013 2h48m06.1s | -6.0 | 100±10 | 96.3 | 55.2 | 101.10±0.53 | 30.90±0.06 | 22.5±0.5 | 19.8±0.6 | 38.2±0.5 |
| 070113b | 7 Jan. 2013 23h03m08.9s | -1.5 | 3.5±0.3 | 89.8 | 51.2 | 102.09±0.06 | 28.03±0.06 | 23.0±0.3 | 20.1±0.3 | 38.3±0.3 |
| 090113 | 9 Jan. 2013 18h41m04.2s | -2.0 | 4.3±0.3 | 94.0 | 62.5 | 102.60±0.36 | 31.89±0.33 | 22.7±0.4 | 19.5±0.4 | 38.4±0.4 |
| 110113 | 11 Jan. 2013 23h56m01.3s | -1.5 | 3.0±0.4 | 90.6 | 64.1 | 109.11±0.46 | 31.51±0.07 | 23.6±0.4 | 20.8±0.5 | 38.2±0.4 |
| 120113 | 11 Jan. 2013 0h08m46.3s | -0.5 | 1.3±0.2 | 93.2 | 72.5 | 104.82±0.04 | 27.46±0.03 | 22.2±0.4 | 19.2±0.5 | 38.2±0.4 |
| 130113 | 13 Jan. 2013 23h09m59.9s | -5.5 | 65±6 | 92.3 | <67.3 | 111.15±0.04 | 30.61±0.02 | 23.6±0.5 | 20.8±0.5 | 38.3±0.5 |
| 180113 | 18 Jan. 2013 22h58m05.2s | -3.0 | 7.8±0.7 | 90.3 | 55.3 | 112.12±0.08 | 31.18±0.03 | 22.1±0.3 | 19.1±0.3 | 38.3±0.3 |

Table 3. Orbital elements (J2000) for the double-station ρ-Geminid meteors discussed in the text, and value of the $D_{SH}$ dissimilarity function. The averaged orbit (for N=10 meteors) is also included.

| Meteor code | a (AU) | e | i (°) | Ω (°) ± $10^{-5}$ | ω (°) | q (AU) | P (yr) | $T_J$ | $D_{SH}$ |
|---|---|---|---|---|---|---|---|---|---|
| 030112 | 2.55±0.20 | 0.746±0.020 | 6.6±0.2 | 281.94757 | 258.56±0.24 | 0.646±0.006 | 4.07 | 2.96±0.10 | 0.10 |
| 060112 | 2.75±0.14 | 0.753±0.013 | 3.3±0.1 | 284.94058 | 253.39±0.16 | 0.681±0.003 | 4.58 | 2.85±0.08 | 0.12 |
| 050113 | 2.52±0.15 | 0.741±0.017 | 2.8±0.1 | 285.66103 | 257.85±0.17 | 0.652±0.004 | 4.00 | 2.99±0.09 | 0.06 |
| 070113a | 2.56±0.19 | 0.734±0.022 | 4.7±0.2 | 286.80531 | 254.30±0.79 | 0.679±0.007 | 4.10 | 2.98±0.12 | 0.10 |
| 070113b | 2.64±0.13 | 0.745±0.013 | 3.1±0.1 | 287.67550 | 257.07±0.09 | 0.671±0.003 | 4.29 | 2.92±0.06 | 0.07 |
| 090113 | 2.67±0.20 | 0.739±0.021 | 5.2±0.2 | 289.51575 | 251.53±0.52 | 0.697±0.004 | 4.37 | 2.90±0.10 | 0.11 |
| 110113 | 2.61±0.18 | 0.750±0.018 | 5.8±0.1 | 291.77423 | 257.38±0.69 | 0.653±0.005 | 4.23 | 2.91±0.09 | 0.08 |
| 120113 | 2.57±0.16 | 0.730±0.018 | 2.7±0.1 | 291.80112 | 252.13±0.06 | 0.695±0.004 | 4.14 | 2.96±0.09 | 0.09 |
| 130113 | 2.62±0.21 | 0.750±0.022 | 5.4±0.1 | 293.78014 | 257.26±0.04 | 0.654±0.005 | 4.24 | 2.90±0.10 | 0.09 |
| 180113 | 2.67±0.13 | 0.733±0.014 | 5.2±0.1 | 298.86338 | 249.59±0.11 | 0.712±0.003 | 4.38 | 2.90±0.08 | 0.12 |
| Average | 2.62±0.15 | 0.742±0.016 | 4.7±0.1 | 289.27646 | 254.71±0.29 | 0.674±0.004 | 4.24 | 2.93±0.10 | 0.07 |

Table 4. Aerodynamic pressure for the flare exhibited by RGE meteors.

| Meteor code | Height (km) | Velocity (km s$^{-1}$) | Aerodynamic pressure (MPa) |
|---|---|---|---|
| 030112 | 62.2±0.5 | 22.6±0.5 | $(1.1±0.1)·10^{-1}$ |
| 060112 | 61.6±0.5 | 21.9±0.5 | $(1.1±0.1)·10^{-1}$ |
| 070113a | 58.2±0.5 | 21.2±0.5 | $(1.6±0.2)·10^{-1}$ |
| 070113b | 56.7±0.5 | 21.9±0.5 | $(2.1±0.2)·10^{-1}$ |
| 090113 | 63.0±0.5 | 21.2±1.0 | $(8.8±1.3)·10^{-2}$ |
| 110113 | 65.0±0.5 | 20.2±0.5 | $(6.1±0.7)·10^{-2}$ |
| 180113 | 60.2±0.5 | 20.6±1.0 | $(1.2±0.2)·10^{-1}$ |

**FIGURES**

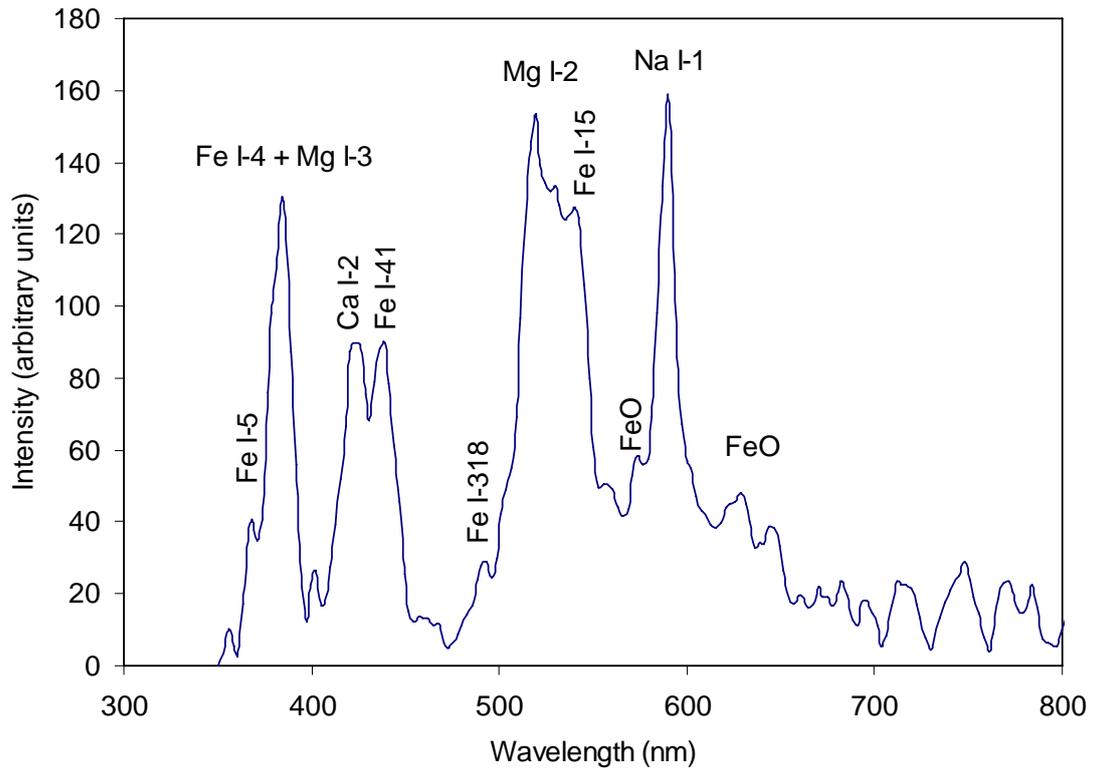

Figure 1. Calibrated emission spectrum integrated along the atmospheric trajectory of the 130113 meteor, where the main lines identified in the signal have been highlighted.

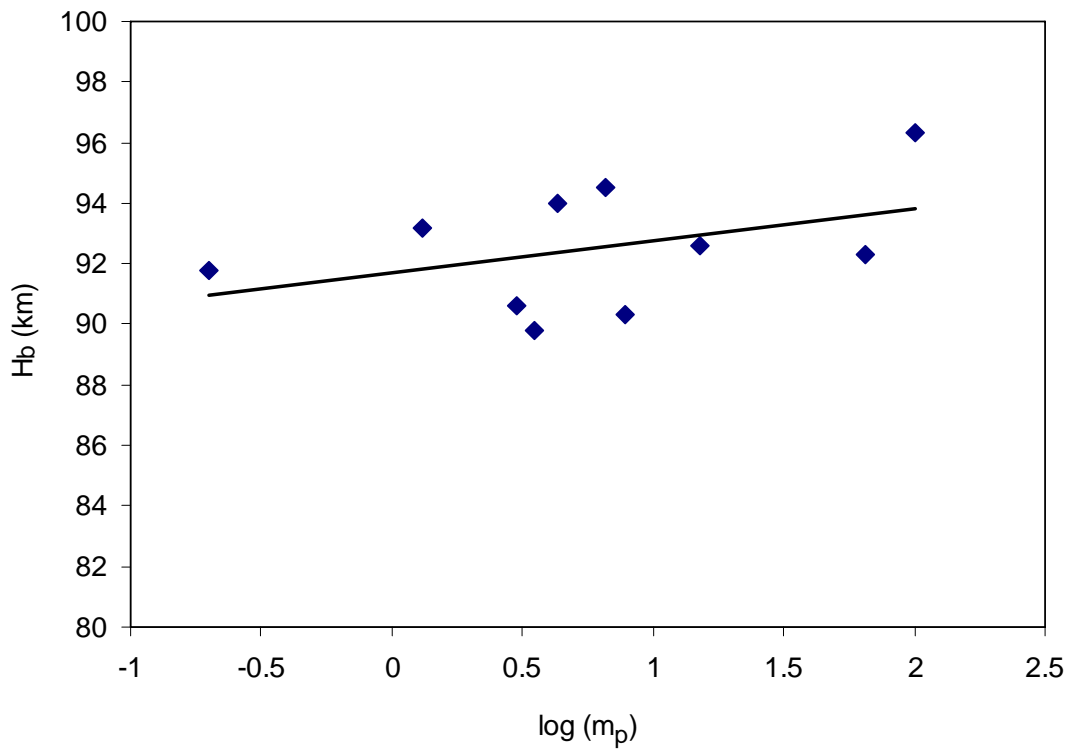

Figure 2. Meteor beginning height $H_b$ vs. logarithm of the photometric mass $m_p$ of the meteoroid. Solid line: linear fit for measured data.

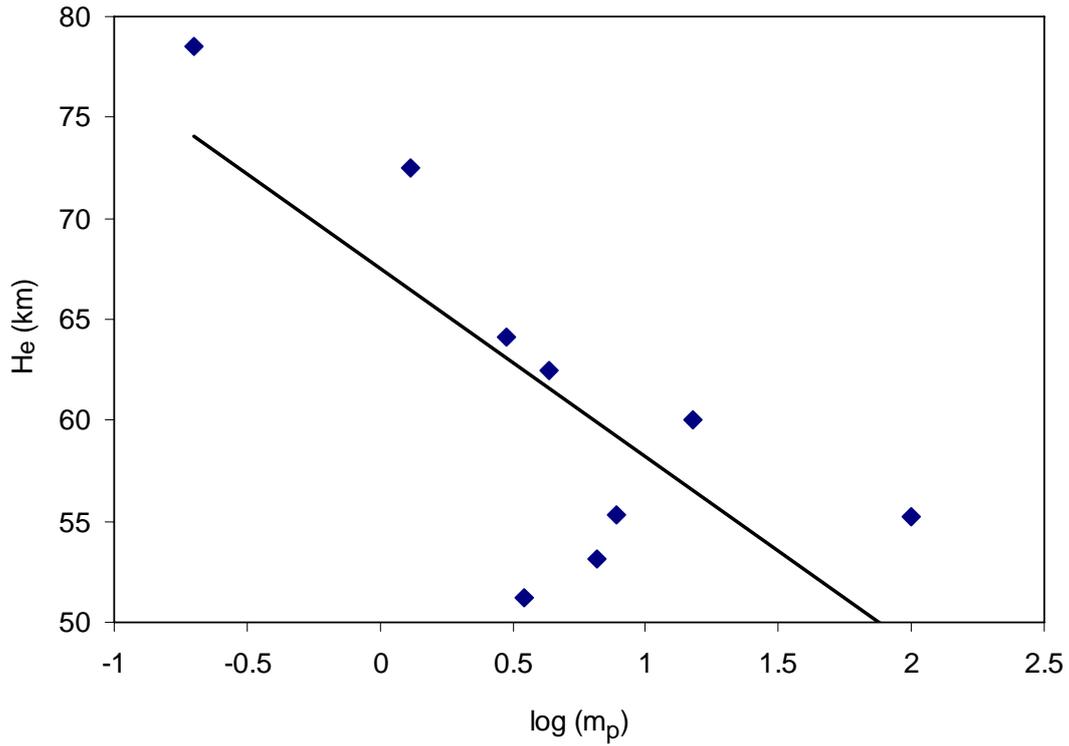

Figure 3. Meteor terminal height $H_e$ vs. logarithm of the photometric mass $m_p$ of the meteoroid. Solid line: linear fit for measured data.

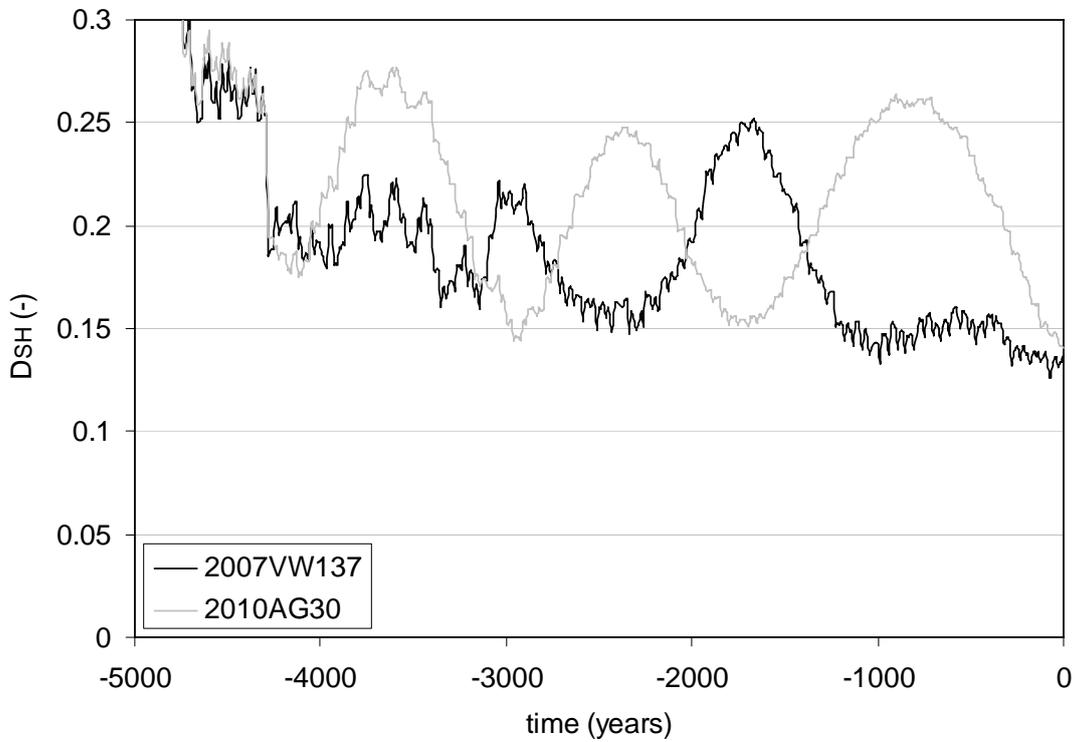

Figure 4. Evolution from present time of the $D_{SH}$ criterion for the RGE meteoroid stream and NEOs 2007VW137 and 2010AG30.

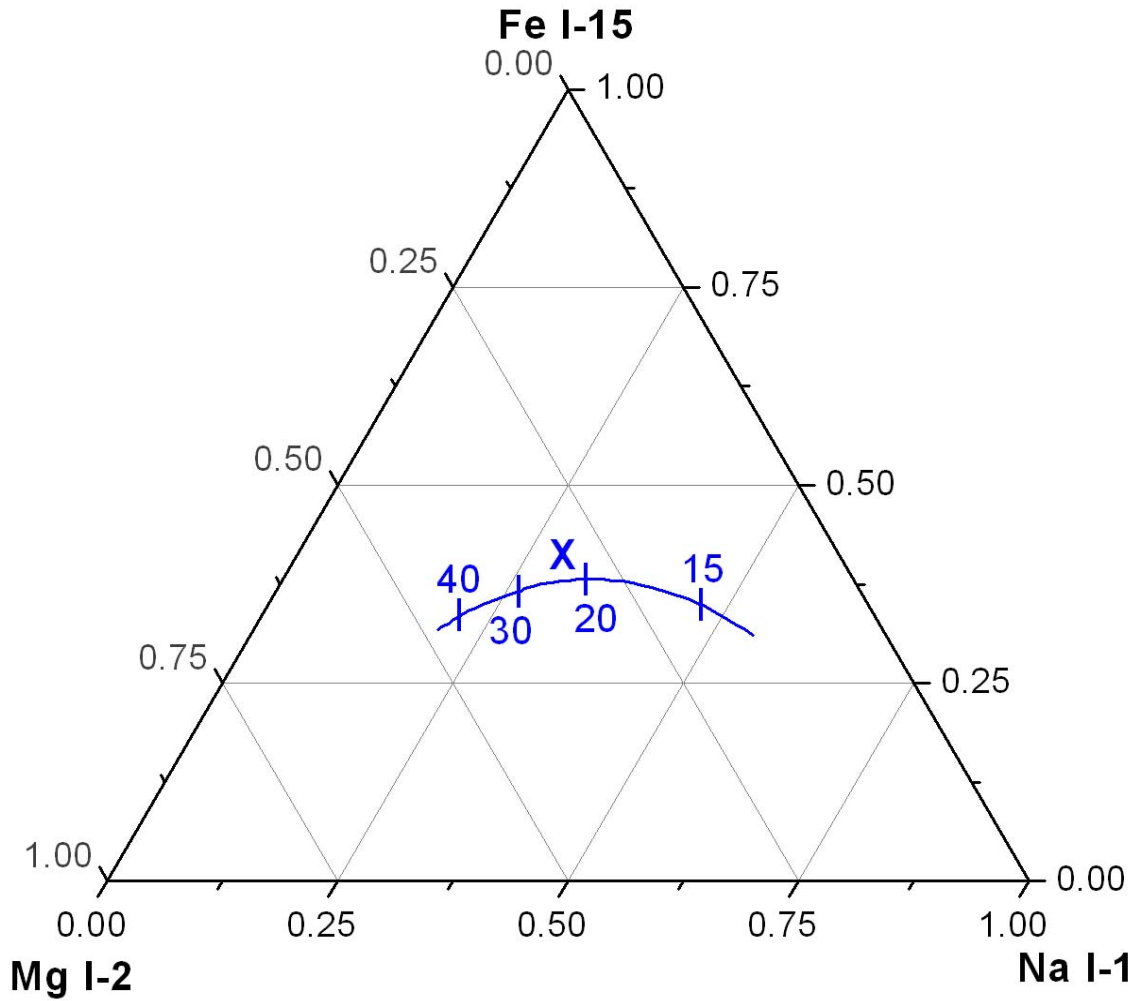

Figure 5. Expected relative intensity (solid line), as a function of meteor velocity (in km s$^{-1}$), of the Na I-1, Mg I-2 and Fe I-15 multiplets for chondritic meteoroids (Borovička et al., 2005). The cross shows the experimental relative intensity obtained for the 130113 meteor.